\documentclass[seceqn]{elsart}

\usepackage{amsmath,latexsym,amssymb,tables,axodraw,psfig,bbm,graphicx,eufrak}
\usepackage{macros,slashed,wick}
\usepackage[dvips]{epsfig}

\input eqalign.sty

\begin{document}

\begin{frontmatter}

\title{Chiral Ward identities, automatic O($a$) improvement and the gradient flow}

\author[FZJ]{Andrea Shindler}
\address[FZJ]{IAS, IKP and JCHP, Forschungszentrum J\"ulich, 52428 J\"ulich, Germany}

\maketitle
\begin{abstract}
Non-singlet chiral Ward identities for fermionic operators at positive flow-time
are derived using standard techniques based on local chiral variations of the action and of local operators. 
The gradient flow formalism is applied to twisted mass fermions 
and it is shown that automatic O($a$) improvement for Wilson twisted mass 
fermions at maximal twist is a property valid also at positive flow-time. 
A definition of the chiral condensate that 
is multiplicatively renormalizable and automatically O($a$) improved
is then derived.
\end{abstract}

}

\end{frontmatter}
\cleardoublepage

\section{Introduction}
\label{sec:intro}
The gradient flow (GF), for gauge~\cite{Luscher:2010iy} 
and fermion~\cite{Luscher:2013cpa} fields, used in combination with a lattice regulator, 
can probe the non-perturbative dynamics of QCD
in advantageous manners.
It is defined by a differential equation that gauge and fermion
fields satisfy as a function of the space-time coordinates $x$ and of a new
scale, the flow-time $t$.
In a way the flow equations modify the short-distance behavior of local fields and
can be considered as a continuous form of stout smearing~\cite{Morningstar:2003gk}.

Smearing has been a useful tool for lattice gauge theories 
since many years~\cite{Albanese:1987ds,Gusken:1989qx,Morningstar:2003gk}.
It was never clear though whether smeared operators had a continuum limit
or not. The only exception was the discussion of the continuum limit for smeared Wilson
loops in a paper by Narayanan and Neuberger~\cite{Narayanan:2006rf}.

The works by L\"uscher and Weisz~\cite{Luscher:2010iy,Luscher:2011bx,Luscher:2013cpa}
give us a complete understanding of the continuum limit
of observables at non-vanishing flow-time. This allows us to use the GF
to define observables that otherwise would be difficult
to compute with standard methods.
For example the GF has been used to give operational definitions to the
fundamental parameters of QCD as the strong coupling~\cite{Luscher:2010iy,Fodor:2012td,Fritzsch:2013je}
or other quantities, as the chiral condensate~\cite{Luscher:2013cpa}
and the energy-momentum tensor~\cite{Suzuki:2013gza,DelDebbio:2013zaa}.
The GF can also be used to define new relative ways to set the scale in 
lattice QCD calculations~\cite{Luscher:2010iy,Borsanyi:2012zs} or 
to define a renormalization scheme for fermionic operators~\cite{Monahan:2013lwa}.

To study correlation functions involving fields
at positive flow-time it is, in some cases, advantageous to consider a suitable extra-dimensional local field theory
where the flow-time $t$ is the extra-dimensional coordinate and 
QCD is located at the $t=0$ boundary.
The theory contains the fundamental fields and appropriate Lagrange multipliers
to guarantee that the fundamental fields, once the Lagrange
multipliers are integrated out, satisfy the proper GF equations at positive flow-time.

In this work we derive non-singlet chiral Ward identities
considering the transformation properties of the action and of local operators under a local
chiral variation. We derive axial and vector Ward identities when the local 
variation is performed at the $t=0$ boundary 
and bulk Ward identities that defines $4+1$ conserved axial and vector currents.
To obtain the Ward identities we use two different perspectives.

The first one takes into account the dependence of the GF equations solutions 
on the initial conditions. A local chiral variation of the initial conditions 
propagates into a variation of local fields located in the bulk. 
Using the Lagrange multipliers it is possible to obtain a universal 
formula that relates the chiral variation located at the boundaries
with a multilocal correlation function that depends exclusively on the 
type of symmetry transformation. This generates additional terms to the standard Ward identities.
The same idea has been investigated in~\cite{DelDebbio:2013zaa}
in the context of space-time symmetries.

The second is to consider fermion fields at positive flow-time
simply as integration variables of the extra-dimensional theory.
We extend the $SU(N_f)_L \times SU(N_f)_R$ chiral symmetry transformations 
for the fermion fields and the corresponding Lagrange multipliers at positive flow-time.
It becomes clear then that the QCD chiral Ward identities get modified by boundary terms.
These boundary terms are identical to the additional terms obtained with the previous method.
This second approach provides a general framework to determine Ward identities
in the bulk and at the boundary of the $4+1$ theory
that can be easily extended to any operator one is interested in. 
As a check we compare the chiral Ward identities with the identities obtained by L\"uscher
in~\cite{Luscher:2013cpa} finding perfect agreement.

Chiral symmetry transformations for the $(4+1)$-dimensional theory have several applications. 
In this work, after extending the GF formalism for twisted mass fermions,
we prove automatic O($a$) improvement for correlators computed with Wilson twisted mass fermions at maximal twist
at non-zero flow-time.
As a by-product we show that the newly proposed way to determine
the chiral condensate with standard Wilson fermions~\cite{Luscher:2013cpa}
can be easily generalized to Wilson twisted mass
with the additional advantage that there is no need
to determine any improvement coefficient.

\section{Gradient flow}
\label{sec:gf}
In this section, to fix the notation, we give a short introduction to 
the GF for gauge and fermions fields~\cite{Luscher:2010iy,Luscher:2013cpa}. 
We consider the $SU(N)$ gauge potential, $G_{\mu}(x)$
and the quark and anti-quark fields $\psi(x)$ and $\psibar(x)$.
The Dirac, color and flavor indices, unless specifically needed,
are suppressed for simplicity.

The Yang-Mills gradient flow~\cite{Luscher:2010iy} of gauge fields as a function 
of the flow-time $t$ is defined as follows
\be
\partial_t B_{\mu}=D_{\nu,t}G_{\nu\mu}\,,
\label{eq:GF_gauge}
\ee
where
\be
G_{\mu\nu}=\partial_{\mu}B_{\nu}-\partial_{\nu}B_{\mu}+[B_{\mu},B_{\nu}]\,,
\qquad
D_{\mu,t}=\partial_{\mu}+[B_{\mu},\,\cdot\;]\,,
\ee
and the initial condition on the flow-time-dependent field $B_{\mu}(t,x)$
at $t=0$ is  given by the fundamental gauge field.
The flow-time $t$ has a time-squared dimension, 
but in the following, with a slight abuse of notation, 
we will refer to it as a single extra-dimension.

The flow for the quark fields considered in ref.~\cite{Luscher:2013cpa} 
is given by
\be
\partial_t\chi_t=\Delta\chi_t \qquad \partial_t\chibar_t=\chibar_t\overleftarrow{\Delta}\,,
\label{eq:flow_ferm}
\ee
\be
\Delta=D_{\mu,t}D_{\mu,t} \qquad D_{\mu,t}=\partial_{\mu}+B_{\mu}\,,
\label{eq:Dmu_def}
\ee
\be
\overleftarrow{\Delta}=\overleftarrow{D}_{\mu,t}\overleftarrow{D}_{\mu,t} 
\qquad \overleftarrow{D}_{\mu,t}=\overleftarrow{\partial}_{\mu}-B_{\mu}\,,
\label{eq:Dmu_def}
\ee
which, together with the initial conditions
\be
\left.\chi_t\right|_{t=0}=\psi \qquad \left.\chibar_t\right|_{t=0}=\psibar\,,
\label{eq:bc}
\ee
define the time-dependent quark and antiquark 
fields $\chi_t(x)$ and $\chibar_t(x)$.
In the following we will indicate with $\chi_t(x)$ and $\chibar_t(x)$
the solutions of the GF equations and with $\chi(t,x)$ and $\chibar(t,x)$
generic fermion fields functions of $(t,x)$.
We note that the GF equations for the fermion fields contain the gauge fields evolved with
eq.~\eqref{eq:GF_gauge}.

Solving the GF equations at tree-level it becomes clear the smoothing 
effect 
\be
\chi_t(x) = \int d^4y~K(t;x-y) \psi(y)\,,\qquad K(t;x) = \frac{\e^{-\frac{x^2}{4t}}}{\left(4\pi t\right)^{2}}\,,
\ee
with smearing radius of $\sqrt{8t}$.

Using the solutions of the flow equations we can construct 
operators located at positive flow-time and consider correlation functions between them. 
Correlation functions of operators located at positive flow-time can be shown to be 
equivalent to correlation functions defined in a local field theory in $4+1$ dimensions, 
the extra dimension being the flow-time~\cite{ZinnJustin:1986eq,ZinnJustin:1987ux,Luscher:2010iy,Luscher:2013cpa}.
The fields content of the $4+1$ field theory consists of 
the gauge and fermion fields at vanishing flow-time, at positive flow-time
and the corresponding Lagrange multipliers. 
The Lagrange multipliers are needed to make sure that
once they are integrated over the gauge and fermion fields
satisfy the appropriate flow equations.
The local extra-dimensional theory is a useful theoretical tool to investigate
correlators involving operators at positive flow-time. For example
the renormalizability of the pure-gauge gradient flow has been established~\cite{Luscher:2011bx}
to all orders of perturbation theory.

In this work we consider only the fermionic sector 
of the theory and details about the gauge sector, the ghost sector and the gauge fixing terms
can be found in refs.~\cite{Luscher:2010iy,Luscher:2013cpa}.
The fermionic part of the action reads
\be
S_{\rm F} = S_{\rm F,QCD} + S_{\rm F,fl}\,,
\ee
where $S_{\rm F,QCD}$ is the usual fermionic QCD action 
including the quark and gauge fields at zero flow-time
\be
S_{\rm F,QCD} = \int d^4x~\psibar(x)\left[ \gamma_\mu D_\mu + M\right]\psi(x)\,,
\ee
where $M$ is the mass matrix of the theory and $D_\mu = \partial_\mu + G_\mu$. 
The bulk action $S_{\rm F,fl}$ of the $(4+1)$-dimensional theory including the fields at non-zero flow-time 
is given by
\be
S_{\rm F,fl} =\int_0^{\infty}\rmd t\int\rmd^4x
  \left[\lambdabar(t,x)
         \left(\partial_t-\Delta\right)\chi(t,x) +\chibar(t,x)
  \left(\overleftarrow{\partial}_t-\overleftarrow{\Delta}\right)
  \lambda(t,x)\right]\,,
\label{eq:action_bulk}
\ee
where $\lambda(t,x)$ and $\lambdabar(t,x)$ are the Lagrange multiplier
for the fermion fields and have energy-dimension of $\left(4+1\right)/2$.
The relative sign between the two terms in eq.~\eqref{eq:action_bulk}
is dictated by charge conjugation symmetry defined in app.~\ref{sec:appB}.
Integrating out the Lagrange multipliers one identifies correlation functions of fields 
evolved with their flow equations to correlation functions in
a local field theory in $4+1$ dimensions, the extra dimension being the flow-time. 

If we consider a perturbative expansion, like for example in dimensional regularization,
a gauge fixing term should be added to the action.
In the following we will consider Ward identities (WIs)
involving gauge invariant operators, thus all possible additional
contributions coming from the gauge-fixing terms vanishes
and we can safely discard them. Additionally we will always have in mind
a non-perturbative definition of the theory using a lattice regulator 
where gauge-fixing terms are not needed.

\section{Chiral symmetry}
\label{sec:chiral}

A global $SU_L(N_f) \times SU_R(N_f)$ chiral symmetry transformation 
\be
\begin{cases}
\psi(x) \rightarrow  {\rm exp}\left\{i\left(\alpha_V^a \frac{T^a}{2} + \alpha_A^a \frac{T^a}{2}\gamma_5\right)\right\} \psi(x) \vspace{0.5cm}\\
\psibar(x) \rightarrow  \psibar(x) {\rm exp}\left\{i\left(-\alpha_V^a \frac{T^a}{2} + \alpha_A^a \frac{T^a}{2}\gamma_5\right)\right\}
\label{eq:chiral}
\end{cases}
\ee
leaves the massless QCD continuum action invariant.
$T^a/2$ are the generators of the SU($N_f$) flavor group in the fundamental 
representation (e.g. Pauli matrices for $N_f = 2$, Gell-Mann matrices for $N_f = 3$ 
with $a=1 \ldots N_f^2-1$) and they satisfy
\be
\Tr\left[T^a T^b\right] = 2 \delta^{ab}\,, \quad
\left[\frac{T^a}{2},\frac{T^b}{2}\right] = i f^{abc} \frac{T^c}{2}\,, \quad
\left\{\frac{T^a}{2},\frac{T^b}{2}\right\} = i d^{abc} \frac{T^c}{2} + \frac{\delta^{ab}}{2} \mathbbm{1}\,,
\ee
where $f^{abc}$ are the SU($N_f$) structure constants, $d^{abc}$ are symmetric coefficients.
Using standard methods considering an infinitesimal
local chiral transformation
\be
\begin{cases}
\delta \psi(x) = \left[i \alpha_V^a(x) \frac{T^a}{2} + i \alpha_A^a(x) \frac{T^a}{2} \gamma_5\right]\psi(x)\,, \\
\delta \psibar(x) = \psibar(x)\left[-i \alpha_V^a(x) \frac{T^a}{2} + i \alpha_A^a(x) \frac{T^a}{2} \gamma_5\right]\,, \\
\end{cases}
\label{eq:deltapsi}
\ee
the corresponding Noether currents, the so-called axial and vector currents,
can be determined
\be
A_\mu^a(x) = \psibar(x)\gamma_\mu\gamma_5\frac{T^a}{2}\psi(x)\,, \qquad 
V_\mu^a(x) = \psibar(x)\gamma_\mu\frac{T^a}{2}\psi(x)\,,
\ee
and the variations of the action reads
\be
i\frac{\delta S_{\rm QCD}}{\delta \alpha_A^a(x)} = \partial_\mu A^a_\mu(x) - \psibar(x) \gamma_5 \left\{\frac{T^a}{2},M\right\}\psi(x)\,,
\ee
\be
i\frac{\delta S_{\rm QCD}}{\delta \alpha_V^a(x)} = \partial_\mu V^a_\mu(x) + \psibar(x) \left[\frac{T^a}{2},M\right]\psi(x) \,.
\ee
For a generic observable $\mathcal{O}$ the variation under the transformations~\eqref{eq:deltapsi} is
\be
i\frac{\delta \mcO }{\delta \alpha_A^a(x)} = 
-\left\{\psibar(x)\left[\gamma_5 \frac{T^a}{2}\right] \frac{\delta \mcO}{\delta \psibar(x)} - 
\frac{\delta \mcO}{\delta \psi(x)}\left[\gamma_5 \frac{T^a}{2}\right] \psi(x)\right\} \,.
\ee
\be
i\frac{\delta \mcO }{\delta \alpha_V^a(x)} = 
\left\{\psibar(x)\left[\frac{T^a}{2}\right] \frac{\delta \mcO}{\delta \psibar(x)} + 
\frac{\delta \mcO}{\delta \psi(x)}\left[\frac{T^a}{2}\right] \psi(x)\right\} \,.
\ee

\subsection{Operators at positive flow-times}

In this section we want to derive non-singlet chiral WIs when the probe operator $\mcO$  
depends on quark and gauge fields solutions of the GF equations~(\ref{eq:GF_gauge},\ref{eq:flow_ferm}).
We derive the variations of the action and the probe operator in the classical theory.
We then consider an interacting theory where a sensible regulator, 
as the lattice or dimensional regularization, is used.
Once the local fields are properly renormalized and the regulator is removed we expect
the same WIs to be valid for the renormalized fields~\cite{Bochicchio:1985xa}. 
 
To determine the form of the chiral WIs we need to understand how the
operator $\mcO$ transforms under a local chiral transformation.
A local chiral transformation of 
fermion fields at vanishing flow-time modifies the initial conditions
of the flow-equations.
Consequently the fermion fields solutions of the flow equations
at positive flow-time are altered. In other words the 
transformation properties under chiral symmetry of the fermion fields at vanishing
flow-time are propagated to the fermion fields at non-vanishing flow-time.

These variations can be determined considering the functional dependence of the fermion fields
at positive flow-time from the initial conditions of the flow equations, i.e.
evaluating the following expression
\bea
i\frac{\delta \mcO }{\delta \alpha_A^a(x)} = 
&-&\int_0^{\infty} dt~\int d^4z~\left\{\left[\psibar(x)\gamma_5 \frac{T^a}{2}\right] \overline{J}(t,z;0,x)
\frac{\delta \mcO}{\delta \chibar_t(z)}  \right. \nonumber \\
&+&\left. \left[\gamma_5 \frac{T^a}{2} \psi(x)\right] J(t,z;0,x)  \frac{\delta \mcO}{\delta \chi_t(z)}\right\} \,.
\label{eq:delta_Ot}
\eea
where the Jacobians are defined as
\be
J(t,z;s,x) = \theta(t-s) \frac{\delta \chi_t(z)}{\delta \chi_s(x)}\,, \qquad
\overline{J}(t,z;s,x) = \theta(t-s) \frac{\delta \chibar_t(z)}{\delta \chibar_s(x)}\,.
\label{eq:jaco}
\ee
Eq.~\eqref{eq:delta_Ot} is a non-local expression due to the presence of the Jacobian matrices
$J$ and $\overline{J}$, thus it would
be rather difficult to study its divergences and renormalization properties.
This problem can be solved if the operator $\mcO$ does not depend on the Lagrange multipliers $\lambda$ and $\lambdabar$.
In fact in this specific case it is possible to evaluate the expression on the r.h.s of eq.~\eqref{eq:delta_Ot}
obtaining
\be
\left\langle i\left[\frac{\delta \mcO}{\delta \alpha_A^a(x)}\right]_R \right\rangle = 
- \left\langle \mcO_R \widetilde{P}_R^a(0,x)\right\rangle\,,
\label{eq:delta_Ot2}
\ee
where
\be
\tilde{P}^a(t,x) = \lambdabar(t,x)\frac{T^a}{2}\gamma_5\chi(t,x)
  +\chibar(t,x)\frac{T^a}{2}\gamma_5\lambda(t,x)\,,
\ee
and 
\be
\tilde{P}_R^a = Z_{\tilde{P}}\tilde{P}^a\,.
\ee

The renormalization of $Z_{\tilde{P}}$, discussed in ref.~\cite{Luscher:2013cpa},
is multiplicative and it turns out to be equal to one even if the interacting theory 
is regularized with Wilson fermions.
If the operator $\mcO$ contains only fermion and gauge fields at positive flow-times 
and does not contain Lagrange multipliers it renormalizes multiplicatively~\cite{Luscher:2013cpa}
depending on the total number $n$ of fermion and anti-fermion fields, i.e. 
\be
\mcO_R = \left[Z_\chi\right]^{n/2} \mcO\,,
\ee
where $Z_\chi$ is the normalization factor for the fermion fields and Lagrange multipliers at positive flow-time
$\chi_R = Z_\chi^{1/2}\chi$, $\lambda_R = Z_\chi^{-1/2}\lambda$. 
 
Eq.~\eqref{eq:delta_Ot2} is a specific case of a more general result
(see app.~\ref{sec:appC} for the derivation) that we can formally write as
\bea
\frac{1}{\mcZ_{\chi,\lambda}}\int\mcD\left[\chi,\chibar \right] \mcD\left[\lambda,\lambdabar \right] 
\mcO \left[\lambdabar(s,x) \Gamma^a(s,x) \right.  &+&  \left. 
\overline{\Gamma^a}(s,x)\lambda(s,x) \right] \exp\left\{-S_{F,fl} \right\} = \nonumber \\
&=& \int_s^{\infty} dt~\int d^4z~\left[\overline{\Gamma^a}(s,x) \overline{J}(t,z;s,x) \frac{\delta \mcO}{\delta \chibar(t,z)}  
\right. \nonumber \\
&+&\left. \Gamma^a(s,x)J(t,z;s,x)\frac{\delta \mcO}{\delta \chi(t,z)}\right]_0\,.
\label{eq:delta_chiral1}
\eea
where 
\be
\mcZ_{\chi,\lambda} = \int\mcD\left[\chi,\chibar \right] \mcD\left[\lambda,\lambdabar \right] \exp\left\{-S_{F,fl} \right\}\,,
\ee
and the index $0$ indicates that the r.h.s of eq.~\eqref{eq:delta_chiral1} has to be evaluated
with fields $\chi$ and $\chibar$ solutions of the GF equations.

Eq.~\eqref{eq:delta_chiral1} is useful because it relates the symmetry transformation of $\mcO$, induced
by the boundary conditions, to a multi-local correlation function whose form is solely dictated by the type of transformation,
here parametrized by the functions $\Gamma$ and $\overline{\Gamma}$. 
Additionally the equation is valid if $\mcO$, $\Gamma^a$ and $\overline{\Gamma}^a$ do not depend
on the Lagrange multipliers $\lambda$ and $\lambdabar$.

We are now in the position to derive the chiral WIs using standard techniques obtaining
\be
\left\langle \left[\partial_\mu A^a_{\mu,R}(x) - P^{\{a,M_R\}}_R(x) +
\widetilde{P}_R^a(0,x)\right] \mcO_R\right\rangle = 0\,,
\label{eq:PCACt}
\ee
\be
\left\langle \left[\partial_\mu V^a_{\mu,R}(x) + S^{\left[a,M_R\right]}_R(x)  + \overline{S}_R^a(0,x)\right] \mcO_R\right\rangle = 0\,,
\label{eq:CVCt}
\ee
where
\be
\overline{S}^a_R = Z_{\overline{S}} \overline{S}^a\,,
\ee
and
\be
\overline{S}^a(t,x) = \lambdabar(t,x)\frac{T^a}{2}\chi(t,x)-\chibar(t,x)\frac{T^a}{2}\lambda(t,x)\,.
\ee
The renormalization of $\overline{S}^a$ following the same argumentation used for $\widetilde{P}^a$
is also multiplicative.
For a generic mass matrix the scalar and pseudoscalar densities read
\be
P^{\{a,M_R\}}_R(x) = Z_P \psibar(x) \gamma_5 \left\{\frac{T^a}{2},M_R\right\}\psi(x)\,,
\ee
and
\be
S^{\left[a,M_R\right]}_R(x) = Z_S \psibar(x) \left[\frac{T^a}{2},M_R\right]\psi(x)\,.
\ee

Eq.~\eqref{eq:PCACt} was already derived in ref.~\cite{Luscher:2013cpa}
using the defining equations for the fermion propagators and Wick contractions.
It is interesting to see that the same relation can be obtained using standard 
symmetry methods. 
This derivation also elucidates the origin of the additional term proportional to $\widetilde{P}^a$,
i.e. it is the variation of the probe operator at positive flow-time 
resulting from the chiral transformation of the initial conditions
of the flow equations.

Eq.~\eqref{eq:CVCt} is new and is the vector WI with an external probe located
at positive flow-time. In this case the variation of the external operator can be worked out
using the general formula~\eqref{eq:delta_chiral1} obtaining that it is proportional
to $\overline{S}^a$.

\section{Chiral symmetry at positive flow-time}
\label{sec:WI}
A different way to derive chiral WIs at positive flow-time is to extend the chiral
symmetry transformations~\eqref{eq:chiral} to the $(4+1)$-dimensional theory.
This extension is more general and it allows us to find the corresponding 
$(4+1)$-dimensional Noether currents and the extension of the WIs for positive 
flow-time.

The chiral symmetry transformation of the fermion fields at vanishing flow-time
are the standard ones for a Dirac field. The boundary conditions~\eqref{eq:bc} 
extend in a natural way the $SU_L(N_f) \times SU_R(N_f)$ chiral symmetry transformations to the 
fermion fields and non-zero flow-time, i.e.
\be
\begin{cases}
\chi(t,x) \rightarrow  {\rm exp}\left\{i\left(\alpha_V^a \frac{T^a}{2} + \alpha_A^a \frac{T^a}{2}\gamma_5\right)\right\} \chi(t,x) \vspace{0.5cm}\\
\chibar(t,x) \rightarrow  \chibar(t,x) {\rm exp}\left\{i\left(-\alpha_V^a \frac{T^a}{2} + \alpha_A^a \frac{T^a}{2}\gamma_5\right)\right\}\,.
\label{eq:chiral_chi}
\end{cases}
\ee

Since the flow equations~\eqref{eq:flow_ferm} together with the initial 
conditions~\eqref{eq:bc} are invariant under the chiral transformations~\eqref{eq:chiral}, 
the time-dependent fields transform in the same way under chiral transformations as
the fields at zero flow-time. For example the scalar and pseudoscalar densities
\be
S^{a}(t,x)=\chibar(t,x)\frac{T^a}{2}\chi(t,x)\,,\qquad P^{b}(t,x)=\chibar(t,x)\gamma_5\frac{T^a}{2}\chi(t,x)\,,
\ee
transform in the same way as the corresponding composite fields at vanishing flow-time~\cite{Luscher:2013vga}.

We want to ask the question how the Lagrange multiplier fields $\lambda$ and
$\lambdabar$ transform under chiral symmetry.
The invariance of the flow equation for fermions can be seen as a consequence
of the invariance under chiral transformation of the bulk action $S_{\rm F,fl}$~\eqref{eq:action_bulk}.
This implies that the chiral transformations are
\be
\begin{cases}
\lambda(t,x) \rightarrow  {\rm exp}\left\{i\left(\alpha_V^a \frac{T^a}{2} - \alpha_A^a \frac{T^a}{2}\gamma_5\right)\right\} \lambda(t,x) \vspace{0.5cm}\\
\lambdabar(t,x) \rightarrow  \lambdabar(t,x) {\rm exp}\left\{i\left(-\alpha_V^a \frac{T^a}{2} - \alpha_A^a \frac{T^a}{2}\gamma_5\right)\right\}\,.
\label{eq:chiral_lambda}
\end{cases}
\ee 
We note the opposite sign for the axial transformation coefficient with respect to
the standard transformation. 
Another way to understand that these are the correct chiral transformations of the Lagrange multipliers
is to consider the bulk action $S_{\rm F,fl}$ and note that the Lagrange multipliers $\lambda$ and $\lambdabar$
are equivalent to the differential operators
\be
\lambdabar \rightarrow \frac{\delta}{\delta \mcF}\,, \qquad 
\lambda \rightarrow \frac{\delta}{\delta \mcFbar}\,,
\ee
where
\be
\mcF(t,x) = \left[\partial_t - \Delta\right]\chi(t,x)\,, \qquad 
\mcFbar(t,x) = \chibar(t,x)\left[\overleftarrow{\partial}_t-\overleftarrow{\Delta}\right]\,.
\ee

We use the chiral transformations for the fermions fields~\eqref{eq:chiral_chi} and 
for the Lagrange multipliers~\eqref{eq:chiral_lambda} to derive chiral
WIs.
Using standard techniques we consider an infinitesimal local
SU($N_f$) $\times$ SU($N_f$) chiral transformations of the fermionic fields 
\be
\begin{cases}
\delta \chi(t,x) = \left[i \alpha_V^a(t,x) \frac{T^a}{2} + i \alpha_A^a(t,x) \frac{T^a}{2} \gamma_5\right]\chi(t,x)\,, \\
\delta \chibar(t,x) = \chibar(t,x)\left[-i \alpha_V^a(t,x) \frac{T^a}{2} + i \alpha_A^a(t,x) \frac{T^a}{2} \gamma_5\right]\,, \\
\delta \lambda(t,x) = \left[i \alpha_V^a(t,x) \frac{T^a}{2} - i \alpha_A^a(t,x) \frac{T^a}{2} \gamma_5\right]\lambda(t,x)\,, \\
\delta \lambdabar(t,x) = \lambdabar(t,x)\left[-i \alpha_V^a(t,x) \frac{T^a}{2} - i \alpha_A^a(t,x) \frac{T^a}{2} \gamma_5\right]\,. \\
\end{cases}
\label{eq:delta_chiral}
\ee
It is interesting to note that the densities $\widetilde{P}^a$ and $\overline{S}^a$ 
form a multiplet as can be seen performing an infinitesimal axial and vector rotation
\be
\delta_A \widetilde{P}^b = f^{abc} \alpha_A^a\overline{S}^c\,, \quad \delta_A \overline{S}^b = f^{abc} \alpha_A^a \widetilde{P}^c\,,
\ee
\be
\delta_V \widetilde{P}^b = f^{abc} \alpha_A^a\widetilde{P}^c\,, \quad \delta_V \overline{S}^b = f^{abc} \alpha_A^a \overline{S}^c\,.
\ee

To take into account properly boundary effects we discretize the
flow-time direction with spacing $\epsilon$ and 
we treat separately the cases $t=0$ and $t>0$.
Discretizing the flow-time direction the bulk action reads
\be
S_{\rm F,fl} = \epsilon \sum_{t=0}^{T-\epsilon}\int\rmd^4x
  \left[\lambdabar(t,x)
         \left(\partial_t-\Delta\right)\chi(t,x) +\chibar(t,x)
  \left(\overleftarrow{\partial}_t-\overleftarrow{\Delta}\right)
  \lambda(t,x)\right]\,,
\label{eq:action_bulk_lattice}
\ee
where 
\be
\partial_t \chi(t,x) = \frac{1}{\epsilon} [ \chi(t+\epsilon,x) - \chi(t,x) ]\,,
\label{eq:partialmu}
\ee
\be
\chibar(t,x) \overleftarrow{\partial}_t = \frac{1}{\epsilon} [ \chibar(t+\epsilon,x) - \chibar(t,x) ]\,.
\ee
The $\Delta$ operator is the lattice Laplacian
\be
\Delta = \nabla_{\mu,t} \nabla_{\mu,t}^*\,, \qquad 
\overleftarrow{\Delta} = \overleftarrow{\nabla}_{\mu,t} \overleftarrow{\nabla}_{\mu,t}^*\,,
\ee
where $\nabla_\mu$ and $\nabla_\mu^*$ are the gauge-covariant forward and backward difference operators 
in presence of the flow-time dependent gauge field.

Computing the local chiral variation of the action at $t=0$ we obtain the so-called PCAC relation
\be
\partial_\mu A^a_\mu(x) - P^{\{a,M\}}(x) + \widetilde{P}^a(0,x)=0 \,.
\label{eq:AWI0}
\ee
Once again we find an additional term with respect to the standard PCAC. 
In the approach we are following in this section 
the additional term is the result of the chiral variation of the boundary terms.

We consider now a generic operator
\be
\mcO = \phi_1(t_1,x_1)\cdots\phi_n(t_n,x_n)\,,
\ee
where $\phi_1,\cdots,\phi_n$ can be fermion or gauge fields. The WIs in the interacting theory,
valid for the renormalized fields once the regulator has been removed, is identical to eq.~\eqref{eq:PCACt}.
Following the line of reasoning of the previous section one might wonder why we have not considered
a variation of the operator $\mcO$. The reason exemplifies the differences between the two methods.
In this second approach the fermion fields at positive flow-time are in general {\it not} solutions
of the GF equations but only integration variables. As such there is no complicated dependence
of these fermion fields from other fermion fields located at smaller flow-time.
In the previous section the fermion fields were solutions of the GF equations, i.e.
the integration over the Lagrange multipliers was already performed, thus the non-local dependence on the
initial conditions.
The generalization of the chiral symmetry transformations for the $(4+1)$-dimensional theory
gives us another tool to obtain eq.~\eqref{eq:PCACt}. 

Using the transformations in~\eqref{eq:delta_chiral} we can derive bulk chiral WIs.
If we perform a local axial variation at a flow-time $s>0$ and 
we assume that $s<t_i$ for all the $t_i$ arguments of $\mcO$ the axial WI after $\epsilon \rightarrow 0$
is
\be
\left\langle \left[ \partial_s \tilde{P}^a(s,x) + \partial_\mu \mcA_\mu^a(s,x) \right]\mcO_R\left(\{t_0\}\right)\right\rangle = 0 \qquad s>0\,, \quad s < \left\{t_0\right\} 
\label{eq:AWIt}
\ee
where we have denoted generically with $\{t_0\}=\left\{t_1,\ldots,t_N\right\}$ the flow-time
dependence of the operator $\mcO$. The current $\mcA_\mu^a$ is
\be
\mcA_\mu^a(s,x) = \lambdabar(s,x)\frac{T^a}{2}\gamma_5~\left(\lD_{\mu,t} -D_{\mu,t}\right)\chi(s,x)
  -\chibar(s,x)\frac{T^a}{2}\gamma_5\left(\lD_{\mu,t} - D_{\mu,t}\right)\lambda(s,x)\,.
\ee
Even if we use Wilson fermions as a lattice regulator at the boundary, the chiral symmetry of the bulk action implies that 
the WI~\eqref{eq:AWIt} is an exact relation in the bare theory. 
Since we expect the same WI to be valid after renormalizing the fields and removing the regulator
we can conclude that the normalization factors of $\tilde{P}^a(s,x)$ and $\mcA_\mu^a(s,x)$ are equal to one
for all positive flow-times and the regulator can be safely removed.
We recall that operators located at positive flow-time are renormalized with their
field content and because $Z_\chi = Z_\lambda^{-1}$ 
the WI~\eqref{eq:AWIt} for $s>0$ is consistent with this result.

Introducing a $4+1$ axial current $\mathcal{A}_M$ with $M=0,\ldots,4$
where $0$ denotes the flow-time direction
\be
\mathcal{A}_M^a = (\tilde{P}^a,\mcA_\mu^a)\,,
\ee
we can rewrite the WI for $s>0$ in a suggestive way 
\be
\left\langle \partial_M \mcA_M^a(s,x) \mcO_R\left(\{t_0\}\right)\right\rangle = 0\,.
\label{eq:AWI_Dp1}
\ee
where
\be
\partial_M = (\partial_t,\partial_\mu)\,.
\ee
We note a dimensional mismatch between the different components of the $(4+1)$-dimensional axial current.
This is a direct consequence of the different dimensionality between the Lagrange multipliers 
and the fermion fields that is related to the different dimensionality between the 
flow-time and any space-time dimension.

Integration of eq.~\eqref{eq:AWIt} over the 4-dimensional space-time leads to
\be
\int d^4 x\left\langle \partial_s \widetilde{P}^a(s,x) \mcO_R(\{t_0\})\right\rangle = 0\,,
\label{eq:ptilde_ind}
\ee
This result can also be obtained using the GF equations and observing that
\be
\int d^4x~K(t,y;s,x)~K(s,x;0,v) = K(t,y;0,v)\,,
\ee
with $t>s>0$. 
Eq.~\eqref{eq:ptilde_ind} is interesting because if we consider the short flow-time expansion
\be
\widetilde{P}^a(s,x) = c_{\tilde{P}}(s) \widetilde{P}^a(0,x) + {\rm O}(s)\,,
\label{eq:sde_ptilde}
\ee
eq.~\eqref{eq:ptilde_ind}, satisfied order by order in $s$, 
immediately implies that $c_{\tilde{P}}$ is a constant and because of the renormalization
group equations applied to eq.~\eqref{eq:sde_ptilde} also independent of the coupling.
A tree-level calculation thus shows that $c_{\tilde{P}}(s)=1$.

We integrate now the WI along the flow-time direction from $\epsilon$ and $t$ with $0<\epsilon<t<t_i$ 
and we obtain
\be
\left\langle \left[\mcA_0^a(\epsilon,x) - \mcA_0^a(t,x)\right] \mcO_R\left(\{t_0\}\right)\right\rangle = 
\left\langle \left[\partial_\mu \int_\epsilon^t ds~\mcA_\mu^a(s,x)\right] \mcO_R\left(\{t_0\}\right)\right\rangle\,.
\ee
To perform the limit $\epsilon\rightarrow 0$  we need to make sure that the integral
in the r.h.s has, at most, integrable singularities.
One needs to classify all the operators of dimension $5$ or less that can mix with $\mcA_\mu$.
The first thing we notice is that the operators appearing in the short flow-time
expansion of $\mcA^a_\mu$ should contain at least one Lagrange multiplier.
This because otherwise any two-point function between such operators 
and $\mcA^a_\mu$ would vanish.
This implies that there can be operators of dimension
$4$ or $5$, but chiral symmetry, extended to the Lagrange multipliers, charge conjugation and O($4$) symmetry
exclude them. We are left with at most a logarithmic divergence when $s\rightarrow 0$
and this divergence is integrable.

Using eq.~\eqref{eq:PCACt} we can then write
\be
-\left\langle \widetilde{P}^a(t,x) \mcO_R\left(\{t_0\}\right)\right\rangle = 
\left\langle \left[\partial_\mu \overline{A}_{\mu,R}^a(t,x) - P^{\{a,M\}}_R(x) \right] \mcO_R\left(\{t_0\}\right)\right\rangle\,,
\label{eq:intPCAC}
\ee
where 
\be
\overline{A}_{\mu,R}^a(t,x) = Z_A A_\mu^a(x) + \int_0^t ds~\mcA_\mu(s,x)\,.
\label{eq:A_5D}
\ee
We can interpret eq.~\eqref{eq:intPCAC} as a generalization of the axial WI
at positive flow-time, where the l.h.s. of eq.~\eqref{eq:intPCAC}, following the general result of eq.~\eqref{eq:delta_chiral1},
is the variation of the operator $\mcO$ induced by a chiral transformation
performed on the quark fields at a flow-time $t<t_i$.

We show now with an example the usefulness of this approach 
deriving the identity obtained by L\"uscher that 
allows to relate the chiral condensate to pseudoscalar correlation functions (cf. with eq. (4.9) of ref.~\cite{Luscher:2013cpa}).
We consider the WI in eq.~\eqref{eq:AWI_Dp1} with an operator $\mcO$
function of fermion fields localized at flow-time
$t$. We do not assume any ordering on the flow-times thus we need to add a contact term to the WI when $t=s$
stemming from the local variation of the operator
\be
\left\langle \left[\partial_s \tilde{P}^a(s,x) + \partial_\mu \mcA_\mu^a(s,x)\right]
\mathcal{O}_R(t,y)\right\rangle = \left\langle i \left[\frac{\delta \mathcal{O}(t,y)}{\delta \alpha_A^a(s,x)}\right]_R\right\rangle \,.
\label{eq:WI2_quant}
\ee
The r.h.s. can be computed in a standard manner if we consider
$\mcO(t,y) = P^b(t,y)$
\be
\langle 
\left[\partial_s \tilde{P}^a(s,x) + \partial_\mu \mcA_\mu^a(s,x) \right] P_R^b(t,y)\rangle = 
- \langle S_R^{\left\{a,b\right\}}(s,y) \rangle 
\delta(t-s) \delta (x-y)\,,
\ee
where
\be
S^{\left\{a,b\right\}}(s,y) = \chibar(s,y)\left\{\frac{T^a}{2},\frac{T^b}{2}\right\}\chi(s,y)\,.
\ee
We recall that $P^a(t,y)$ and $S^a(t,y)$ renormalize multiplicatively with $Z_\chi$~\cite{Luscher:2013cpa}.
Integrating this WI over the whole $4$-dimensional space-time volume we obtain
\be
\int d^4x~\langle 
\partial_s \tilde{P}^a(s,x) P_R^b(t,y)\rangle = 
- \langle S_R^{\left\{a,b\right\}}(s,y) \rangle \delta(t-s)\,.
\ee
Having included the contact term we can now integrate in the flow-time
in the range $\epsilon < s < T$ with $T>t$ and perform the limit $\epsilon \rightarrow 0$
\be
\lim_{\epsilon \rightarrow 0}\int d^4x~\int_\epsilon^T ds~\langle 
\partial_s \tilde{P}^a(s,x) P_R^b(t,y)\rangle = 
- \lim_{\epsilon \rightarrow 0} \int_\epsilon^T ds~\langle S_R^{\left\{a,b\right\}}(s,y) \rangle \delta(t-s)\,.
\ee
The boundary term at $T$ does not contribute because propagators $\wick{1}{<*\lambda(T,x) >*{\chibar}(t,x)}$ 
and $\wick{1}{<*\chi(t,x) >*{\lambdabar}(T,x)}$ vanish for $T>t$~\cite{Luscher:2013cpa}.
We thus obtain
\be
\int d^4x~\langle \tilde{P}^a(0,x) P_R^b(t,y)\rangle = \langle S_R^{\left\{a,b\right\}}(t,y) \rangle\,.
\label{eq:ptilde_cond}
\ee
This is exactly the relation obtained in ref.~\cite{Luscher:2013cpa},
that allows the determination of the chiral condensate (cfr. eq. 4.15 of ref.~\cite{Luscher:2013cpa}).
This example shows that it is possible to use standard techniques based
on chiral symmetry transformations to derive WIs
involving fields at the boundary and solutions of the GF
for positive flow-time.

We remark that in the derivation of eq.~\eqref{eq:ptilde_cond} it is not necessary to send $\epsilon \rightarrow 0$.
If we integrate over the flow-time between $s_0$ and $T$ with $s_0<t<T$ we obtain
\be
\int d^4x~\langle 
\tilde{P}^a(s_0,x) P_R^b(t,y)\rangle = \langle S_R^{\left\{a,b\right\}}(t,y) \rangle\,.
\ee
This result is independent of $s_0$ as far as $0<s_0<t$ as can be seen from eq.~\eqref{eq:ptilde_ind}.
It is only to make contact with the physics of pions that it is necessary to place $\tilde{P}^a$
at the boundaries.

\subsection{Vector Ward identities}

Using the procedure outlined in the previous section we can derive also the so-called PCVC relation at the boundary 
$t=0$ and the corresponding relation in the bulk
\be
\begin{cases}
\partial_\mu V^a_\mu(x) + \psibar(x) \left[\frac{T^a}{2},M\right]\psi(x) + \overline{S}^a(0,x)=0  & t=0 \\
\partial_t \overline{S}^a(t,x) + \partial_\mu \mcV_\mu^a(t,x)=0  & t > 0 
\end{cases}
\label{eq:WI1}
\ee
where
\be
\overline{S}^a(t,x) = \lambdabar(t,x)\frac{T^a}{2}\chi(t,x)-\chibar(t,x)\frac{T^a}{2}\lambda(t,x)\,,
\ee
and
\be
\mcV^a_\mu(t,x) = \lambdabar(t,x)\frac{T^a}{2}\left(\lD_{\mu,t} - D_{\mu,t}\right)\chi(t,x) +
\chibar(t,x)\left(\lD_{\mu,t} - D_{\mu,t}\right)\frac{T^a}{2}\lambda(t,x)\,.
\ee

As for the axial case we can rewrite the second relation in~\eqref{eq:WI1} for $t>0$ 
introducing a $4+1$ vector current $\mathcal{V}_M$ with $M=0,\ldots,4$
where the $0$ index refers to the flow-time direction
\be
\mathcal{V}_M^a = (\overline{S}^a(t,x),\mcV^a_\mu(t,x))\,.
\ee
For $t>0$ the vector current $\mathcal{V}_M^a$ is conserved
\be
\partial_M \mathcal{V}_M^a=0\,.
\ee
Using as a probe $\mcO\left(\{t_0\}\right)$, the VWI with $t_0>s$ reads
\be
\left\langle \left[\partial_t \mcV_0^a(s,x) + \partial_\mu \mcV_\mu^a(s,x)\right]\mcO_R(\{t_0\})\right\rangle=0\,,
\ee
and the generalization of the standard VWI is
\be
-\left\langle \overline{S}^a(t,x) \mcO(t_0,y)\right\rangle = 
\left\langle \partial_\mu \overline{V}_\mu^a(t,x) \mcO(\{t_0\})\right\rangle\,,
\label{eq:intCVC}
\ee
where 
\be
\overline{V}_\mu^a(t,x) = Z_V V_\mu^a(x) + \int_0^t ds~\mcV_\mu(s,x)\,.
\ee

\section{Twisted mass QCD}
\label{sec:tmQCD}
In this section we consider the GF with twisted mass fermions~\cite{Frezzotti:2000nk}.
For simplicity we consider a flavour doublet of mass degenerate twisted
mass fermions. The extension to non-degenerate twisted mass fermions~\cite{Frezzotti:2003xj} 
does not pose any additional complication.
For reviews about twisted mass fermions see refs.~\cite{Sharpe:2006pu,Sint:2007ug,Shindler:2007vp}.
In the twisted basis the continuum twisted mass action
\be
S_{\rm F,tmQCD} = \int d^4x~\psibar(x)\left[\gamma_\mu D_\mu + m + i \mu_{\rm q}\gamma_5 \tau^3\right]\psi(x)\,,
\ee
is obtained by a chiral rotation of a twist angle $\omega={\rm atan}\left(\mu_q/m\right)$
\be
\begin{cases}
\psi(x) \rightarrow  {\rm exp}\left\{i\left(\omega \frac{\tau^3}{2}\gamma_5\right)\right\} \psi(x) \vspace{0.5cm}\\
\psibar(x) \rightarrow  \psi(x) {\rm exp}\left\{i\left(\omega \frac{\tau^3}{2}\gamma_5\right)\right\}\,.
\end{cases}
\ee 
of the continuum QCD action.
To obtain the fermionic part of the action 
for the $(4+1)$-dimensional theory we perform the same twist rotation
on the fermion fields at positive flow-time
\be
\begin{cases}
\chi(t,x) \rightarrow  {\rm exp}\left\{i\left(\omega \frac{\tau^3}{2}\gamma_5\right)\right\} \chi(t,x) \vspace{0.5cm}\\
\chibar(t,x) \rightarrow  \chibar(t,x) {\rm exp}\left\{i\left(\omega \frac{\tau^3}{2}\gamma_5\right)\right\}\,.
\label{eq:twist}
\end{cases}
\ee
The Lagrange multipliers are also a flavor doublet and the twist rotation 
is obtained using the chiral symmetry transformations defined in eq.~(\ref{eq:chiral},\ref{eq:chiral_lambda})
\be
\begin{cases}
\lambda(t,x) \rightarrow  {\rm exp}\left\{i\left(- \omega \frac{\tau^3}{2}\gamma_5\right)\right\} \lambda(t,x) \vspace{0.5cm}\\
\lambdabar(t,x) \rightarrow  \lambdabar(t,x) {\rm exp}\left\{i\left(- \omega \frac{\tau^3}{2}\gamma_5\right)\right\}\,.
\label{eq:twist_lambda}
\end{cases}
\ee 
The exact chiral symmetry of the bulk term in the $4+1$ theory implies that this term of the action
stays invariant under a twist rotation, i.e. the total twisted mass action is
\be
S_{\rm F} = S_{\rm F,tmQCD} + S_{\rm F,fl}\,,
\label{eq:twisted_action}
\ee
where $S_{\rm F,fl}$ is given by eq.~\eqref{eq:action_bulk}.
This is not surprising because the GF equations~\eqref{eq:flow_ferm} and the initial
conditions~\eqref{eq:bc} are invariant under a twist rotation.

The twist rotations in eqs.~(\ref{eq:twist},\ref{eq:twist_lambda}) define the 
operators at positive flow-time in the twisted basis.
Fermions bilinears formed only by fermion fields or only
Lagrange multipliers transform in a standard manner.
Bilinears made of fermion and Lagrange multiplier fields 
twist in a different way and, as an example, we show how the twist rotation applies to $\tilde{P}^a$ and
$\overline{S}^a$ 
\be
\begin{cases}
\tilde{P}^a \longrightarrow \cos \omega \tilde{P}^a + \sin \omega \overline{S}^a\,, \qquad a=1,2 \\
\tilde{P}^a \longrightarrow \tilde{P}^a\,, \qquad a=0,3\,,
\end{cases}
\label{eq:ptilde_twist}
\ee
\be
\begin{cases}
\overline{S}^a \longrightarrow \cos \omega \overline{S}^a  + \sin \omega \tilde{P}^a\,, \qquad a=1,2 \\
\overline{S}^a \longrightarrow \overline{S}^a\,, \qquad a=0,3\,.
\end{cases}
\label{eq:sbar_twist}
\ee
Using the action in eq.~\eqref{eq:twisted_action} and the chiral 
transformations~(\ref{eq:chiral_chi},\ref{eq:chiral_lambda})
we can derive the PCAC and PCVC relations
\be
\begin{cases}
\partial_\mu A_\mu^a(x) - 2mP^a(x) -i \mu_q \delta^{a3} S^0(x) + \tilde{P}^a(0,x)=0 & t=0  \\
\partial_t \tilde{P}^a(t,x) + \partial_\mu \mcA_\mu^a(t,x)=0 & t>0 \,,
\end{cases}
\label{eq:WI_tm}
\ee
and
\be
\begin{cases}
\partial_\mu V_\mu^a(x) + 2 \mu_q \epsilon^{3ab}P^b(x) + \overline{S}^a(0,x) = 0 & t=0  \\
\partial_t \overline{S}^a(t,x) + \partial_\mu \mcV_\mu^a(t,x) = 0 & t>0 \,.
\end{cases}
\label{eq:VWI_tm}
\ee

The equivalence between QCD and tmQCD in the interacting theory has been proven in~\cite{Frezzotti:2000nk} using as a regulator
overlap fermions. We note that the exact chiral symmetry of the bulk action does not add any complication 
and the result of~\cite{Frezzotti:2000nk} is valid for correlation functions involving
local operators also at positive flow-time.

\section{Wilson twisted mass fermions}
\label{sec:auto}
The GF equations can be applied to a fermion field independently on the
fermionic lattice action of choice. Here we choose to study the interplay 
between the GF and Wilson twisted mass (Wtm) fermions. In particular we want to analyze
the O($a$) cutoff effects of this formulation for operators defined
at positive flow-time and we want to see if it is possible to extend the
property of automatic O($a$) improvement.

Since we will only be concerned with O($a$) effects of the $(4+1)$-dimensional theory
we do not need to specify the exact form of the gauge action.
The form of the unimproved fermionic lattice action at the boundary $t=0$ is 
\be
  S_{\rm Wtm}[\psi,\psibar,U] =a^4\sum_x\psibar(x)\Big[D_{\rm W} + m_0 + i\mu_0\gamma_5\tau^3\Big]\psi(x)\,,
\label{eq:WtmQCD2}
\ee
where $D_{\rm W}$ is the usual Wilson massless operator
\be
D_{\rm W} = \frac{1}{2}\left[\gamma_\mu\left(\nabla_\mu + \nabla_\mu^*\right) - a\nabla^*_\mu\nabla_\mu\right]\,.
\ee
The bulk action is the one defined in~\eqref{eq:action_bulk_lattice} and $m_0$, $\mu_0$ are the bare
untwisted and twisted quark mass.

The long distance properties of Wtm close to the continuum limit
may be described in terms of a local effective theory with action
\be
S_{\rm eff} = S_0+aS_1+a^2S_2+\ldots
\label{eq:eff_action}
\ee
where the leading term, $S_0$, is the action of the target continuum theory~\eqref{eq:twisted_action}
plus the gauge action with properly renormalized parameters.
The higher order terms are linear combination of higher-dimensional operators. For example
\be
S_1 = \int_0^{\infty} dt \int d^4x \sum_i c_i(g_0^2) \mcO_{R,i}(t,x)\,,
\ee
where $\mcO_i(t,x)$ respect the symmetries of the lattice action.
We omit for simplicity the dependence on the renormalization scale.

A correlation function of products of a multiplicatively renormalizable lattice fields, here denoted by $\phi_R=Z_\phi \phi$,
at separated points $(t_i,x_i)$ 
\be
G(t_1,x_1;\ldots;t_n,x_n) = \langle \phi_R(t_1,x_1) \cdots \phi_R(t_n,x_n) \rangle \equiv \langle \Phi_R \rangle
\label{eq:phi}
\ee
takes the form
\be
\left\langle \Phi_R \right\rangle =  \langle \Phi_0 \rangle_0 - a \langle \Phi_0
     S_1 \rangle_0 + a \langle \Phi_1 \rangle_0 + {\rm O}(a^2)\,,
\label{eq:sym_exp1}
\ee
where 
\be
\langle \Phi_0 \rangle_0 \equiv \langle\phi_0(t_1,x_1) \cdots \cdots \phi_0(t_n,x_n) \rangle_0\,,
\ee
\be
\langle \Phi_1 \rangle_0 \equiv \sum_{k=1}^{n}\langle\phi_0(t_1,x_1) \cdots \phi_1(t_k,x_k) \cdots \phi_0(t_n,x_n) \rangle_0\,,
\ee
and $\phi_0,\phi_1$ are renormalized continuum fields. $\phi_1$ is a linear combination of 
local operators of dimension $d_\phi + 1$ that depends on the specific operator $\phi$ and 
are classified according to the lattice symmetries transformation properties of $\phi$.
The expectation values on the right hand side of eq.~\eqref{eq:sym_exp1} are to be taken in the continuum theory
with action $S_0$

In the massless limit Wilson and Wtm fermions have the same lattice action
thus the higher-dimensional operators in the Symanzik effective action are the same.
Using the equations of motion for the quark fields in the chiral limit a possible list of O($a$) terms
is~\cite{Luscher:1996sc,Luscher:2013cpa}
\be
 \qquad {\mathcal O}_1 = 
  i \psibar(x)\sigma_{\mu\nu}F_{\mu\nu}\psi(x)\,, \qquad 
{\mathcal O}_2 =
  \lambdabar(0,x) \lambda(0,x)\,.
\label{eq:sym_op}
\ee

It is important to notice that there are no $t$-dependent terms in the effective action.
The additional operator $\mcO_2$ is a boundary term.
From the point of view of the chiral symmetry transformations~(\ref{eq:chiral_chi},\ref{eq:chiral_lambda})
the absence of O($a$) terms for $t>0$ is a consequence of the chiral invariance of the bulk action.

Additional terms in the effective theory at non-zero quark mass for Wtm fermions
have been classified in~\cite{Luscher:1996sc,Frezzotti:2001ea}. The gradient flow
adds the operators $\mcO_7$ and $\mcO_8$ to the list
\bea
\mcO_3 &=& m \tr \{ F_{\mu\nu}(x) F_{\mu\nu}(x) \}\,,\quad
\mcO_4 = m\psibar(x) \psi(x)\,, \quad 
\mcO_5 = m\mu_{\rm q} i \psibar(x)\gamma_5\tau^3\psi(x)\,, \nonumber \\
\mcO_6 &=& \mu_{\rm q}^2\psibar(x)\psi(x)\,, \quad 
\mcO_7 = m\left(\lambdabar(0,x)\psi(x) + \psibar(x)\lambda(0,x)\right)\,, \nonumber \\ 
\mcO_8 &=& i \mu_{\rm q} \left(\lambdabar(0,x)\gamma_5 \tau^3\psi(x) + \psibar(x)\gamma_5 \tau^3\lambda(0,x)\right)\,.
\label{eq:op_mass}
\eea
The operator $\mcO_7$ is already given in ref.~\cite{Luscher:2013cpa}.
The operator $\mcO_8$ is new and is the one needed to remove O($a \mu_{\rm q}$) effects to the fermion fields and Lagrange multipliers.

\subsection{Automatic O($a$) improvement}

Automatic O($a$) improvement~\cite{Frezzotti:2003ni} is the property of Wtm that physical correlation functions 
with a finite continuum limit made of multiplicatively renormalizable fields are free from O($a$) effects, 
when the lattice parameters are tuned to obtain in the continuum
limit that the renormalized untwisted quark mass vanishes,$m_{\rm R} = 0$.

From the lattice perspective this corresponds to set the bare untwisted mass $m_0$ to its critical
value $\mcr$. The exact way this is achieved is not relevant for what follows, but for 
a discussion and further references on this topic see~\cite{Shindler:2007vp}.

To parametrizes the O($a$) uncertainties 
stemming from the determination of the critical mass, 
i.e. $m_{\rm q} = m_0 - \mcr = $ O($a$) in the Symanzik effective theory
one adds the operator 
\be
\mcO_0 =  \Lambda^2\psibar(x) \psi(x)\,,
\ee
where $\Lambda^2$ is some energy scale squared which depends on the way the critical mass is determined,
e.g. it could be of the order of the QCD scale $\Lambda_{\rm QCD}^2$, or it could 
be something proportional to $\Lambda_{\rm QCD}\mu_{\rm q}$.
This allows to check if potential O($a$) uncertainties in the critical mass contradict 
the property of automatic O($a$) improvement.

The proof of automatic O($a$) improvement for correlation functions containing fields at positive flow-
time follows the proof for the $t=0$ case~\cite{Frezzotti:2005gi,Shindler:2005vj,Aoki:2006nv}. 
We extend the relevant symmetries for the Lagrange multipliers
\be
\mcR^{1,2}_{5,t} \colon
\begin{cases}
\chi(t,x) \rightarrow i \gamma_5 \tau^{1,2} \chi(t,x) \\
\chibar(t,x) \rightarrow  \chibar(t,x) i \gamma_5 \tau^{1,2} \\
\lambda(t,x) \rightarrow -i \gamma_5 \tau^{1,2} \lambda(t,x) \\
\lambdabar(t,x) \rightarrow  - i \lambdabar(t,x) \gamma_5 \tau^{1,2} \,,
\end{cases}
\label{eq:R512_t}
\ee
and
\be
\mcD_t \colon
\begin{cases}
V(t,x;\mu) \rightarrow V^{\dagger}(t,-x-a\hat{\mu};\mu), \\
\chi(t,x) \rightarrow {\rm e}^{3 i \pi/2} \chi(t,-x) \\
\chibar(t,x) \rightarrow  \chibar(t,-x){\rm e}^{3 i \pi/2}\\
\lambda(t,x) \rightarrow {\rm e}^{5 i \pi/2} \lambda(t,-x) \\
\lambdabar(t,x) \rightarrow  \lambdabar(t,-x){\rm e}^{5 i \pi/2}\,.
\end{cases}
\ee
The equivalent transformations for continuum fields,
that with abuse of notation we indicate in the same way, are the same for the fermion
fields, while for the gauge fields the $\mcD_t$ transformation is
$B_\mu(t,x) \rightarrow -B_\mu(t,-x)$.
In the $\mcD_t$ transformation we do not reverse the sign in the $t$ direction
because the flow-time has a space-time dimension of $2$. Thus any derivative 
with respect to the flow-time contained in a generic operator will not change the
parity related to its dimensionality.
To include the twisted mass in the counting of the dimensions of the 
operators appearing in the lattice and continuum Lagrangian one introduces the spurionic
symmetry
\be
\widetilde{\mathcal{D}_t} =  \mathcal{D}_t\times [\mu_{\rm q} \rightarrow -\mu_{\rm q}]\,.
\ee

The lattice action~\eqref{eq:WtmQCD2} at the boundary, and~\eqref{eq:action_bulk_lattice}
in the bulk is invariant under the $\mcR^{1,2}_{5,t} \times \widetilde{\mathcal{D}_t}$ transformation.
If the target continuum theory has a vanishing renormalized untwisted mass, $m_R=0$, it  
is invariant separately under the $\mcR^{1,2}_{5,t}$ and the $\widetilde{\mathcal{D}_t}$
transformations. 
This immediately implies that all the higher-dimensional operators in the Symanzik expansion
contributing to $S_1$ are odd under $\mcR^{1,2}_{5,t}$. This is not true for operators proportional
to the untwisted quark mass, but the maximal twist condition $m_R=0$ forces them to contribute
to higher orders in the Symanzik expansion.
For example the new terms $\mcO_2$ and $\mcO_8$,
containing at least one Lagrange multiplier are odd under $\mcR^{1,2}_{5,t}$, thus they 
vanish once inserted in $\mcR^{1,2}_{5,t}$ even correlation functions.
The same argument applies for the higher-dimensional operators
appearing in the effective theory representations of local operators such as axial currents
or pseudoscalar densities at positive flow-time.

Let us see with specific examples how this works.
The list of operators contributing at O($a$) to the axial current $A_\mu^a$ in the Symanzik effective theory 
is given by
\be
\left(\delta A^a_\mu\right)_1 = \partial_\mu P^a(0,x)\,,\quad  \left(\delta A^a_\mu\right)_2=\tilde{A}_\mu^a(0,x)\,,\nonumber 
\ee
\be
\left(\delta A^a_\mu\right)_3 =  m A_\mu^a(x)\,,\quad 
\left(\delta A^a_\mu\right)_4 =  \mu_{\rm q} \epsilon^{3ab}V_\mu^b(x)\,,
\ee
where 
\be
\tilde{A}_\mu^a(0,x) = \lambdabar(0,x) \gamma_\mu \gamma_5 \frac{T^a}{2} \psi(x) + \psibar(x) \gamma_\mu\gamma_5 \frac{T^a}{2} \lambda(0,x)\,.
\ee
It is easy to check that all these operators have opposite $\mcR_{5,t}^{1,2}$ parity with respect to $A_\mu^a$.
The only exception being the term proportional to the untwisted mass $m$ that vanishes at maximal twist.
Consistently also the $\widetilde{\mcD}_t$ parity is different between $A_\mu^a$ and the higher-dimensional 
operators. This means that their contribution to the effective correlator
vanishes if the continuum limit of the lattice correlator containing $A_\mu^a$ is $\mcR_{5,t}^{1,2}$ even.

Another example is given by the local operator $\tilde{P}^a$. The list of operators contributing at 
O($a$) in the Symanzik effective theory is given by
\be
\left(\delta \tilde{P}^a\right)_1 = \partial_\mu \tilde{A}_\mu^a(0,x)\,, \quad \left(\delta \tilde{P}^a\right)_2 = \hat{P}^a(0,x)\,, \nonumber
\ee
\be
\left(\delta \tilde{P}^a\right)_3 = m \tilde{P}^a(0,x)\,,\quad \left(\delta \tilde{P}^a\right)_4 = \mu_{\rm q} \epsilon^{3ab}\overline{S}^b(0,x)\,,
\ee
where 
\be
\hat{P}^a(0,x) = \lambdabar(0,x) \gamma_5 \frac{T^a}{2} \lambda(0,x)\,.
\ee
As before all the higher-dimensional operators have opposite $\mcR_{5,t}^{1,2}$ parity with respect to $\tilde{P}^a$,
excluding the term proportional to $m$ that anyhow vanishes at maximal twist.

\subsection{Chiral condensate}

As an application of the WIs we have derived, following the strategy suggested by L\"uscher in~\cite{Luscher:2013cpa},
we show how to compute the chiral condensate with Wtm fermions at maximal twist, retaining
the property of multiplicative renormalization and absence of mixing with lower-dimensional operators.
Additionally with Wtm at maximal twist one does not need to compute any improvement coefficient but having leading O($a^2$) effects
discretization errors. 

In the basis we have chosen for the fermion fields, the so-called twisted basis,
the quark condensate at positive flow-time is related to the expectation value of $P^{3}$
\be
\Sigma_t = \frac{\Sigma_t^{uu} + \Sigma_t^{dd}}{2} = - i \langle P^3(t,x) \rangle\,,
\ee
that renormalizes multiplicatively with $Z_\chi$.
To relate the $\Sigma_t$ to the physics of pions we use the VWIs at vanishing and positive flow-time~\eqref{eq:VWI_tm}
and as external probe operator at positive flow-time we choose a pseudoscalar density, i.e
\be
\begin{cases}
\left\langle \left[\partial_\mu V_{\mu,R}^1(0,x) + 2 \mu_R P_R^2(0,x) + \overline{S}_R^1(0,x)\right]P_R^2(t,y)\right\rangle = 0 \\
\left\langle \left[\partial_s \overline{S}^1(s,x) + \partial_\mu \mcV_\mu^1(s,x)\right] P_R^2(t,y) \right\rangle = 
i \langle P^3(t,y) \rangle \delta(t-s) \delta(x-y) \,.
\end{cases}
\ee
Following the same strategy we have used in sec.~\ref{sec:WI} we integrate the first VWI over space-time and 
the second one over space-time and over the flow-time direction from $0$ to $T>t$.
We then obtain 
\be
2 \mu_R \int d^Dx \langle P_R^2(0,x) P_R^2(t,y)\rangle = i \langle P_R^3(t,y) \rangle\,.
\label{eq:int_VWI_tm}
\ee
This equation relates the charged pion correlation function with the chiral condensate 
at positive flow-time. The existence of a conserved vector current guarantees that we can choose the
normalization of twisted mass such that $\mu_R P_R^2$ is renormalization group invariant. This insures that 
eq.~\eqref{eq:int_VWI_tm} is valid with bare correlators up to discretization errors. 
Automatic O($a$) improvement guarantees that the leading discretization errors are of O($a^2$).
Using the standard VWI with an external probe at vanishing flow-time 
\be
\left\langle \left[\partial_\mu V_{\mu,R}^1(0,x) + 2 \mu_R P_R^2(0,x) \right]P_R^2(0,y)\right\rangle = 0\,, \quad x \neq y \\
\ee
and considering the spectral decomposition of the charged pseudoscalar correlators we obtain
for the renormalized chiral condensate 
\be
\Sigma_{\rm R} = \lim_{\mu_R \rightarrow 0}Z_P \frac{\Sigma_t G_\pi}{G_{\pi,t}}\,,
\ee
where $G_\pi$ and $G_{\pi,t}$ are the vacuum-to-pion matrix elements of the 
charged pseudoscalar density respectively at vanishing and positive flow-time.
Given that the renormalization between $\Sigma_t$ and $G_{\pi,t}$ simplifies 
the only renormalization constant needed to compute the chiral condensate is $Z_P$.
It is not surprising that we reproduce the same equation obtained by L\"uscher in~\cite{Luscher:2013cpa}. 
The additional advantage of using Wtm at maximal twist is automatic O($a$) improvement
and the fact that there is no need to compute any improvement coefficient.
In fact the expectation value $i \langle P^3(t,x) \rangle$
is even under $\mcR_{5,t}^{1,2}$ as the charged pseudoscalar correlators thus 
the chiral condensate not only renormalizes multiplicatively 
but it is also not affected by any O($a$) cutoff effects.

With standard or clover Wilson fermions to remove completely O($a$) cutoff effects
it is necessary to add an additional counterterm to the lattice action proportional to 
$\lambdabar \lambda$. The tunable coefficient of this term has been labeled in~\cite{Luscher:2013cpa}
$c_{\rm fl}$. We can write the quark condensate lattice correlator with the insertion of this term
\bea
\langle \ubar(t,x) \gamma_5 u(t,x) &-& \dbar(t,x) \gamma_5 d(t,x)  \rangle = \nonumber \\
&-& a^8 \sum_{y,z} \langle \Tr \left\{ K(t,x;0,y) \gamma_5\left[S_u(y,z) -a c_{\rm fl} \delta_{y,z}\right]K(t,x;0,z)^\dagger \right\}  \nonumber \\
&-& \Tr \left\{ K(t,x;0,y) \gamma_5\left[S_d(y,z) -a c_{\rm fl} \delta_{y,z}\right]K(t,x;0,z)^\dagger \right\}\rangle_{\rm G}\,,
\eea
where $\langle \rangle_{\rm G}$ indicates the gauge average including the fermionic determinant.
The terms proportional to $c_{\rm fl}$ are independent of the flavor and they cancel out.
This is just a reflection of the different $\mcR_{5,t}^{1,2}$ parities between the observable and the O($a$) term.

\section{Concluding remarks}
\label{sec:conclu}

The gradient flow is becoming an important tool to probe
in interesting ways the non-perturbative dynamic of QCD, one aspect being
the non-perturbative renormalization of local fields. The example of the chiral condensate
shows that mixing with lower-dimensional operators can be avoided if we 
relate the quark condensate defined at positive and zero flow-time.
A key role for this relation to hold is played by chiral Ward identities. 

In this work we have shown that it is possible to use local chiral variations
to derive the corresponding Ward identities with operators located at positive flow-time
reproducing the identities obtained by L\"uscher~\cite{Luscher:2013cpa}.
While this alternative derivation is interesting on its own, it is also useful when considering more complicated
operators. In fact Ward identities can be used to give non-perturbative definitions
to the coefficients appearing in small flow-time expansions.

We have then considered the gradient flow in combination with twisted mass fermions.
We have shown that the property of automatic O($a$) improvement for Wilson twisted mass
fermions at maximal twist is still valid at positive flow-time.
Using the vector Ward identity derived for twisted mass fermions we find a definition 
for the chiral condensate that not only is multiplicatively renormalizable but
also automatic O($a$) improved.

\section*{Acknowledgments}

I wish to thank Martin L\"uscher and Agostino Patella for enlightening
discussions on the gradient flow. I want to thank the CERN Theory division,
where part of this work has been done, for their kindness and hospitality.
I also thank Tom Luu for reading a first version of this work and 
for constant encouragement.

\vspace{0.4cm}
\begin{appendix}
\section{Charge conjugation symmetry}
\label{sec:appB}

The charge conjugation symmetry transformation takes the same form for the fields 
at vanishing flow-time, at positive flow time and the Lagrange multipliers.
Here we indicate generically the fermion field as $\phi = \left\{\psi(0,x),\chi(t,x),\lambda(t,x)\right\}$
and the corresponding $\overline{\phi}$. 
For charge conjugation we have
\be
\mathcal{C} \colon
\begin{cases}
   V(t,x;\mu) \longrightarrow  V(t,x;\mu)^*, \\
   \phi    \longrightarrow  
                C^{-1}\overline{\phi}^T,\\
   \overline{\phi} \longrightarrow 
               - \phi^T C ,
\end{cases}
\label{eq:chargeconj}
\ee
where $V(t,x;\mu)$ is the gauge link at positive flow-time and $C$ satisfies
\be
- \gamma_\mu^T = C \gamma_\mu  C^{-1} , \qquad \gamma_5 = C \gamma_5  C^{-1}. 
\ee

\section{Derivation of eq.~\eqref{eq:delta_chiral1}}
\label{sec:appC}
In this appendix we derive eq.~\eqref{eq:delta_chiral1} that we rewrite here for convenience
\bea
\frac{1}{\mcZ_{\chi,\lambda}}\int\mcD\left[\chi,\chibar \right] \mcD\left[\lambda,\lambdabar \right] 
\mcO \left[\lambdabar(s,x) \Gamma^a(s,x) \right.  &+&  \left. 
\overline{\Gamma}^a(s,x)\lambda(s,x) \right] \exp\left\{-S_{F,fl} \right\} = \nonumber \\
&=& \int_s^{\infty} dt~\int d^Dz~\left[\overline{\Gamma}^a(s,x) \overline{J}(t,z;s,x) \frac{\delta \mcO}{\delta \chibar(t,z)}  
\right. \nonumber \\
&+&\left. \Gamma^a(s,x)J(t,z;s,x)\frac{\delta \mcO}{\delta \chi(t,z)}\right]_0\,.
\label{eq:delta_chiral_app}
\eea
where 
\be
\mcZ_{\chi,\lambda} = \int\mcD\left[\chi,\chibar \right] \mcD\left[\lambda,\lambdabar \right] \exp\left\{-S_{F,fl} \right\}\,,
\ee
and the index $0$ indicates that the r.h.s of eq.~\eqref{eq:delta_chiral_app} has to be evaluated
with fields $\chi$ and $\chibar$ solutions of the GF equations.

The proof is based on the observation that, given the form
of the bulk action, an operator linear in $\lambda$ or $\lambdabar$,
as in the l.h.s of eq.~\eqref{eq:delta_chiral_app}, can be generated by a suitable change of variable.
The non-singlet nature of the transformation parametrized by the functions $\Gamma^a$ and $\overline{\Gamma}^a$
guarantees that the Jacobian of this change of variable has unit determinant.

We start considering the kernel $K(t,z;s,x)$ defined as solution of
\be
\begin{cases}
\left(\partial_t - \Delta_z\right) K(t,z;s,x) = 0 & t >s \\
\lim_{t \rightarrow s^+}K(t,z;s,x) = \delta(z-x) & \\
K(t,z;s,x) = 0 & t<s\,. \\
\end{cases}
\ee
The Jacobians $J$ and $\overline{J}$ defined in eq.~\eqref{eq:jaco} satisfy the same equations.
This can be checked applying the gradient flow operator to $J$ and $\overline{J}$ and recalling that in the Jacobians the fields
$\chi$ and $\chibar$ are solutions of the gradient flow equations.

We now consider the expectation value of a generic operator $\mcO$ defined at positive flow-time
and function of the fermion or gauge fields. We assume that $\mcO$ is independent of the Lagrange multipliers.
In the functional integral that defines the expectation value we perform
the following change of variable
\be
\delta \chi(t,z) = i \int_0^tds~\int d^4 x \Gamma^a(s,x)K(t,z;s,x)\alpha^a(s,x)\,,
\label{eq:delta_chi_app}
\ee
\be
\delta \chibar(t,z) = i \int_0^tds~\int d^4 x \overline{\Gamma}^a(s,x)K(t,z;s,x)^{\dagger}\alpha^a(s,x)\,. 
\label{eq:delta_chibar_app}
\ee
As we will show below the Jacobian of this change of variable is one, thus the variation
of the expectation value of $\mcO$ should vanish. Similarly to what happens when deriving 
Ward identities this implies that $\langle \delta \mcO \rangle = \langle O \delta S\rangle$.

The variation of the action according to eq.~(\ref{eq:delta_chi_app},\ref{eq:delta_chibar_app}) is given by
\be
i \frac{\delta S}{\alpha^a(t,y)} = - \left[\lambdabar(t,y)\Gamma^a(t,y) + \overline{\Gamma}^a(t,y) \lambda(t,y)\right]\,.
\ee
If we consider now that the kernels $K$ and $K^\dagger$ coincide with the Jacobians in eq.~\eqref{eq:jaco} 
computing the variation of the operators $\left\langle \delta \mcO \right\rangle$ 
we obtain the desired formula~\eqref{eq:delta_chiral_app}.

We are left to show that the Jacobian of the transformation~(\ref{eq:delta_chi_app}), (\ref{eq:delta_chibar_app})
has determinant equal to one.
From the definition~(\ref{eq:delta_chi_app},\ref{eq:delta_chibar_app}) 
and given that we consider a transformation with infinitesimal $\alpha^a$
this is a result of the non-singlet nature of the transformation, 
i.e. of the fact the matrices $\Gamma^a$ and $\overline{\Gamma}^a$ are traceless.

\end{appendix}

\bibliography{gf}      
\bibliographystyle{h-elsevier}    
\end{document}